\documentclass{PoS}

\title{Evaluation of Israel-Stewart parameters in lattice gauge theory}

\ShortTitle{Evaluation of Israel-Stewart parameters}

\author{\speaker{Yasuhiro Kohno}\\
        Department of Physics, Osaka University, Toyonaka 560-0043, Japan\\
        E-mail: \email{kouno@kern.phys.sci.osaka-u.ac.jp}}

\author{Masayuki Asakawa\\
        Department of Physics, Osaka University, Toyonaka 560-0043, Japan\\
        E-mail: \email{yuki@kern.phys.sci.osaka-u.ac.jp}}

\author{Masakiyo Kitazawa\\
        Department of Physics, Osaka University, Toyonaka 560-0043, Japan\\
        E-mail: \email{kitazawa@kern.phys.sci.osaka-u.ac.jp}}
        
\author{Chiho Nonaka\\
        Department of Physics, Nagoya University, Furo-cho, Chikusa-ku, Nagoya 464-8602, Japan\\
        E-mail: \email{nonaka@hken.phys.nagoya-u.ac.jp}}
        
\abstract{Navier-Stokes equations are known as hydrodynamic equations which take account of effects of dissipations.
          There are, however, problems in the relativistic Navier-Stokes equations, i.e. the equations violate causality.
          Israel-Stewart equations, which evade the problems of Navier-Stokes equations by introducing new parameters, 
          such as the relaxation times, have recently been used in describing the space-time evolution of the quark-gluon plasma produced in high energy heavy ion collisions.
          The viscosities and the relaxation times are related to each other by imposing entropy constraints on the system.
          According to Boltzmann-Einstein principle, the probability distribution of the fluctuation in the energy-momentum tensor is related to the entropy of the system.
          Applying this principle to the entropy in Israel-Stewart theory, one can obtain the ratios of the viscosities to the relaxation times.
          We evaluate the ratios of the viscosities to the relaxation times in SU(3) lattice gauge theory.
          }

\FullConference{The XXVII International Symposium on Lattice Field Theory - LAT2009\\
		 July 26-31 2009\\
		 Peking University, Beijing, China}

\begin{document}

\section{Introduction}
The new form of matter created at RHIC (Relativistic Heavy Ion Collider) has provided physicists unexpected findings.
One of the most impressive results is that the new matter, which is referred to as quark-gluon plasma (QGP), 
behaves like perfect fluid near critical temperature $T_\mathrm{c}$.
The space-time evolution of QGP in high energy heavy ion collisions seems to be understood quantitatively by relativistic ideal hydrodynamics.
However, relativistic dissipative hydrodynamics is needed for more 
quantitative description of the real QGP evolution 
since QGP has small but nonzero viscosities.

The simplest relativistic dissipative hydrodynamics is the first 
order theory which is a relativistic extension of Navier-Stokes 
theory \cite{CE,LDL}.
This theory, however, accompanies problems associated with causality 
violation \cite{WAHandLL}.
Instead of the first order theory, second order dissipative hydrodynamics, 
especially Israel-Stewart (IS) theory \cite{WIandJMS}, has been recently 
actively studied.
IS theory avoids the problems by introducing second order terms 
of dissipation, which physically represent relaxation effects to
the solution of the first order theory.
As an inevitable consequence of introducing the second order terms, 
however, IS theory includes many phenomenological parameters 
in addition to the three transport coefficients in the first order theory.
These parameters cannot be determined within hydrodynamics.
Other approaches based on microscopic theories are needed to 
determine the parameters.

``Perfect fluid''-like behavior of QGP indicates that quarks and 
gluons strongly interact with each other in QGP near $T_\mathrm{c}$ 
\cite{AMY}.
The properties of QGP near $T_{\rm c}$ are not within the reach of 
perturbation theory, since perturbative expansion is poorly converging 
at large couplings.
At present, lattice gauge theory is the only systematic approach which 
can calculate physical quantities in such a non-perturbative region.
There have been done several attempts to measure transport coefficients 
with numerical simulations on the lattice \cite{FKandHWW,ANandSS,HM}.
These analyses evaluate transport coefficients using Kubo formulae, 
which relate transport coefficients to the correlation functions of 
the energy-momentum tensors in Minkowsiki space-time.
On the other hand, 
on the lattice one can calculate correlation functions 
in Euclidean space-time.
An analytic continuation from imaginary-time correlation functions to 
real-time ones therefore is required in this strategy.
In previous studies \cite{FKandHWW,ANandSS,HM}, 
nontrivial ans\"atze have been adopted for the form of real-time
spectral function in order to perform this analytic continuation.
Validity of such ans\"atze, however, 
should be carefully examined.

In the present study, we attempt to constrain phenomenological 
parameters in IS theory on the lattice using a method proposed 
in Refs.~\cite{AM,SP}.
In this method, ratios between the viscosities and the relaxation 
times in IS equations are related to static fluctuations of 
stress tensor in equilibrium through Boltzmann-Einstein 
principle.
Since one does not use temporal correlation functions in this method,
one can avoid the difficulty in the analytic continuation.
The main objective of this work is to evaluate these ratios in 
SU(3) gauge theory on the lattice and to reduce the number of 
phenomenological parameters in IS theory.

\section{Israel-Stewart theory}

In this section, we overview relativistic dissipative hydrodynamics, especially Israel-Stewart (IS) theory \cite{WIandJMS,AM2}.
Basic equations of hydrodynamics are local conservation laws of the energy-momentum and the net charge,
\begin{eqnarray}
\partial_{\mu}T^{\mu\nu}=0,
\label{tmunu} \\
\partial_{\mu}N^{\mu}=0. 
\label{nmu}
\end{eqnarray}
Using arbitrary 4-velocity of fluid $u^{\mu}$ normalized as 
$u^{\mu}u_{\mu}=1$ and the projection onto 3-dimensional space 
$\Delta^{\mu\nu}=g^{\mu\nu}-u^{\mu}u^{\nu}$, 
the energy-momentum tensor $T^{\mu\nu}$ and the charge density $N^{\mu}$ 
are decomposed as
\begin{eqnarray}
T^{\mu\nu}
&=&
\epsilon u^{\mu}u^{\nu} - (p+\Pi)\Delta^{\mu\nu}
+W^{\mu}u^{\nu} + W^{\nu}u^{\mu} + \pi^{\mu\nu},\\
N^{\mu}&=&nu^{\mu}+V^{\mu}, 
\end{eqnarray}
where $\epsilon$, $p$, and $n$ are the energy density, pressure, and 
charge density, respectively.
Eckart used the particle flow as 4-velocity $u^{\mu}$,
\begin{equation}
u^{\mu}\equiv \frac{N^{\mu}}{\sqrt{N_{\mu}N^{\mu}}}, \label{uE}
\end{equation}
which is called Eckart (particle) frame, and $V^{\mu}$ vanishes 
in this frame.
Landau-Lifshitz employed the energy flow for $u^{\mu}$,
\begin{equation}
u^{\mu}\equiv \frac{T^{\mu}_{\nu} u^{\nu}}{\sqrt{u^{\alpha} T_{\alpha}^{\beta}T_{\beta \gamma}u^{\gamma}}}.
\end{equation}
This is Landau-Lifshitz (energy) frame, and $W^{\mu\nu}$ 
vanishes in this frame.
While the values of $W^\mu$ and $V^\mu$ depend on the choice 
of the frame,
\begin{eqnarray}
q^\mu = W^{\mu} - \frac{\epsilon +p}{n}V^{\mu},
\end{eqnarray}
i.e. heat flow in the particle frame, is frame independent 
in first order.

The basic idea of Israel \cite{Israel} for a phenomenological 
derivation of second order hydrodynamics is to incorporate the
effects of dissipation into the entropy current $s^\mu$.
Under the hydrodynamic assumption, i.e. that nonequilibrium states 
are characterized by hydrodynamic variables, $T^{\mu\nu}$ and $N^\mu$,
the entropy current should be given by 
$s^\mu = s^\mu(T^{\rho\sigma},N^\rho)$.
Assuming further that one can expand $s^\mu$ for nonequilibrium
states by the power series of $\Pi$, $\pi^{\mu\nu}$, and $q^\mu$, 
the most general form of entropy current at second order in
dissipative terms, $\Pi$, $\pi^{\mu\nu}$, and $q^\mu$, reads,
\begin{equation}
s^{\mu}=\frac{s_{\mathrm{eq}}}{n}N^{\mu}+\frac{q^{\mu}}{T}+Q^{\mu}, 
\label{smu}
\end{equation}
where 
\begin{equation}
Q^{\mu}
=
-\frac{u^{\mu}}{2T}\left(\beta_{0}\Pi^2
-\beta_1q_{\nu}q^{\nu}+\beta_2\pi_{\nu\lambda}\pi^{\nu\lambda} \right)
-\frac{\alpha_0\Pi q^{\mu}}{T}+\frac{\alpha_1\pi^{\mu\nu}q_{\nu}}{T},
\label{sIS}
\end{equation}
represents the second order contribution to entropy, 
and $s_{\mathrm{eq}}$ and $T$ are the entropy density in 
equilibrium and the local temperature of the system, respectively.
Here, $\beta_i$ and $\alpha_i$ are phenomenological coefficients 
and are not determined by the hydrodynamic assumption.

Requiring the second law of thermodynamics, 
$\partial_{\mu}s^{\mu}\geq 0$, one can constrain the macroscopic 
equations that hydrodynamic variables follow.
If we neglect the second order term in $s^\mu$, the
second law and linearity lead to the first order hydrodynamic 
equations including three transport coefficients, shear and bulk 
viscosities $\eta$ and $\zeta$, and heat conductivity $\lambda$.
Similarly, when one recovers the second order terms, $Q^\mu$, 
in Eq.~(\ref{smu}), the second law leads to constraints
including second order terms. 
IS equations are defined so as to satisfy these constraints
\cite{WIandJMS,Israel}.

IS equations include terms $D\Pi$ and $D\pi$
with $D \equiv u_\mu \partial^\mu$ being the 
derivative along $u^\mu$.
These terms give rise to relaxation effects to the solution of 
first order equations, and cure the causality problem of the 
first order theory.
Proportional coefficients of these terms are given by
\begin{eqnarray}
\tau_\Pi = \beta_0 \zeta, \qquad \tau_\pi = 2\beta_2 \eta, 
\label{tbe}
\end{eqnarray}
respectively, which physically represent the time scale of 
the relaxation in each channel and are called relaxation times.
Equation~(\ref{tbe}) means that the ratios between
viscosities and the relaxation times are related to 
proportional coefficients in the entropy current,
$\beta_0$ and $\beta_2$, as 
\begin{eqnarray}
\beta_0=\frac{\tau_{\Pi}}{\zeta},\qquad
\beta_2=\frac{\tau_{\pi}}{2\eta}.
\label{beta} 
\end{eqnarray}

\section{Boltzmann-Einstein principle}

In what follows, we try to evaluate the ratios in Eq.~(\ref{beta})
on the lattice by connecting $\beta_0$ and $\beta_2$ to 
fluctuations of stress tensor using Boltzmann-Einstein (BE) 
principle.

The entropy and the number of microscopic states in equilibrium are related to each other by Boltznmann relation
\begin{equation}
S(\vec{a}) = \ln W(\vec{a}),
\end{equation}
where $\vec{a}= \left\{ a_1, a_2, a_3, ... \right\}$ represents the set 
of state variables of the system, and 
$S(\vec{a})$ and $W(\vec{a})$ are the entropy and the number of 
microscopic states, respectively, in the system specified by 
state variables $\vec{a}$.
To put it another way, the number of microscopic states is given by 
BE principle \cite{AM,SP},
\begin{equation}
W(\vec{a})=\exp[S(\vec{a})]. 
\label{BE}
\end{equation}
Since the probability that one of the state variables $a_i$ 
takes a certain value, $A$, in equilibrium is given by 
\begin{equation}
P(a_i=A) 
= \frac{W(a_i=A)}{\sum W(\vec{a})},
\end{equation}
BE principle means that the probability distribution of 
state variables in equilibrium is governed by the entropy,
provided that the entropy $S(\vec{a})$ is defined.

>From Eqs.~(\ref{smu}) and (\ref{sIS}), the entropy in unit volume $s$ 
in IS theory at rest frame $u^\mu$ is given by
\begin{equation}
s
= u_\mu s^{\mu}
= s_{\rm eq}-\frac{1}{2T}\left(\beta_0 \Pi^2
-\beta_1q_\mu q^\mu +\beta_2\pi_{\mu\nu}\pi^{\mu\nu} \right).
\label{S}
\end{equation}
Notice that with orthogonality relations 
$u_{\mu}q^{\mu} = u_{\mu}\pi^{\mu\nu} = 0$
the terms including $\alpha_i$ in Eq.~(\ref{sIS})
do not appear in Eq.~(\ref{S}).
As a metter of course, the entropy in Eq.~(\ref{S}) is the same
quantity as that appearing in the BE principle.
Substituting Eq.~(\ref{S}) into Eq.~(\ref{BE}), the number of 
microscopic states specified by $\Pi$, $q^\mu$, and $\pi^{\mu\nu}$
in a volume $V$ is given by
\begin{equation}
W(\bar\Pi ,\bar q^\mu ,\bar\pi^{\mu\nu})
\propto 
\exp[-\frac{V}{2T}\left( \beta_0 \bar\Pi^2 
-\beta_1 \bar q_\mu \bar q^\mu 
+\beta_2 \bar\pi_{\mu \nu} \bar\pi^{\mu \nu} \right)], 
\label{W}
\end{equation}
where $\bar\Pi$, $\bar q^\mu$, and $\bar\pi^{\mu\nu}$ are 
spatial average of each quantity in V, 
\begin{eqnarray}
\bar\Pi = \frac1V \int_V d^3x \Pi(\vec{x}),
\label{bar}
\end{eqnarray}
and so forth.
Equation~(\ref{W}) means that fluctuations of $\bar\Pi$, 
$\bar q^\mu$, and $\bar\pi^{\mu\nu}$ are related to $\beta_i$
in entropy current Eq.~(\ref{smu}).
In particular, distribution of each quantity is of Gaussian form;
for example, that of $\pi_{\mu\nu}$ is given by
\begin{equation}
P(\bar\pi^{\mu\nu})
\propto \exp[-\frac{V}{2T}\beta_2\bar\pi_{\mu\nu}\bar\pi^{\mu\nu}], 
\label{dp}
\end{equation}
and the fluctuation reads
\begin{equation}
\langle (\bar\pi^{\mu\nu})^2 \rangle
= \frac{T}{2V\beta_2}, 
\label{pi^2}
\end{equation}
where we have used $\langle \bar\pi^{\mu\nu} \rangle = 0$.

Since $\beta_i$ are related to fluctuations of stress tensor
through Eq.~(\ref{pi^2}), 
one can evaluate the $\eta$ to $\tau_\pi$ ratio
on the lattice by measuring fluctuations 
$\langle (\bar\pi^{\mu\nu})^2 \rangle$
in a quite straightforward way.
Although this analysis cannot determine the values of 
transport coefficients themselves, 
the number of free parameters in IS theory can be reduced.
Since this method does not refer to correlation functions,
one can avoid the procedure of the analytic continuation, 
which is one of the most nontrivial task in the measurement 
of viscosities using Kubo formulae.

\section{Measurement of fluctuations on the lattice}

The basic idea of lattice gauge theory is to discretize the 
four dimensional Euclid space in a gauge invariant way.
The partition function on the lattice in path-integral 
representation is given by
\begin{eqnarray}
Z = \int dU d\bar{\psi} d\psi \ \mathrm{e}^{-S_G(U)-S_F(\bar{\psi} ,\psi ,U)}
  = \int dU \ \mathrm{Det} F[U] \ \mathrm{e}^{-S_G(U)}. \label{pf}
\end{eqnarray}
Here $S_G$ and $S_F$ are the Euclidean gauge and fermion actions on the lattice, respectively.
$U$, $\bar{\psi}$, and $\psi$ correspond to the gauge, anti-fermion, and fermion fields in the continuum limit, respectively.
$\mathrm{Det} F[U]$ represents the fermion determinant.
In a general procedure of lattice calculations, 
the expectation value of a physical quantity $O$ in equilibrium is computed as an average over gauge configurations:
\begin{equation}
\langle O \rangle = \frac{1}{Z} \int dU \ \mathrm{Det} F[U] \ \mathrm{e}^{-S_G(U)} O. \label{ev}
\end{equation}
In this work we analyze SU(3) pure gauge theory where the fermion part
in Eqs.~(\ref{pf}) and (\ref{ev}) is neglected;
\begin{eqnarray}
Z = \int dU \ \mathrm{e}^{-S_G(U)}, \qquad
\langle O \rangle = \frac{1}{Z} \int dU \ \mathrm{e}^{-S_G(U)} O. 
\label{O}
\end{eqnarray}
In this theory, conserved current corresponding to 
$N^\mu$ in Eq.~(\ref{nmu}) does not exist and the 
hydrodynamic equation is solely given by Eq.~(\ref{tmunu}).
Difference between energy and particle flows, $q^\mu$, cannot 
be defined in this theory.

On the lattice, the energy flow does not exist in the rest 
frame of the lattice.
For $\Pi$ and $\pi_{\mu\nu}$ we thus take $\Pi = \frac13 \sum_{i=1}^3 T_{ii}$
and $\pi^{ij} = T_{ij}$ with $1\le i<j \le3$.
Spatial average in Eq.~(\ref{bar}) should be taken for a volume
$V$ on a given time slice. To eliminate possible artificial effects arising from 
periodicity along spatial direction, $V$ should be considerably 
smaller than the lattice volume in numerical simulations.


\section{Numerical Results}

\begin{table}\label{tbl:para}
\begin{center}
\begin{tabular}{l c l c l c l}
\hline
& $\beta = 6/g^2$ & $a[\mathrm{fm}]$ & Size \ ($N_{\sigma}^3 \times N_{\tau}$) & $T/T_{c}$ & $N_\mathrm{{config}}$\\
\hline
case 1 & $6.499$ & $0.049$ & $32^3 \times 6$ & $2.5$ & 10000\\
\hline
case 2 & $6.499$ & $0.049$ & $32^3 \times 10$ & $1.5$ & 10000\\
\hline
case 3 & $6.872$ & $0.031$ & $48^3 \times 10$ & $2.5$ & 10000\\
\hline
\end{tabular}
\end{center}
\caption{Simulation parameters.
$N_{\sigma}$, $N_{\tau}$, and $N_\mathrm{config}$ represent the number of lattice sites in spatial and temporal directions,
and gauge configurations, respectively.
$a$ and $N_\tau$ are related to the temperature $T$ as $a N_{\tau}=T^{-1}$.
Cases 1 and 2 have the same lattice spacing with the different temperature.
Cases 1 and 3 have the same temperature with the different lattice spacing.}
\label{tbl:ltx-tbl2}
\end{table}

\begin{figure}
\begin{center}
\includegraphics*[angle=0,scale=0.9]{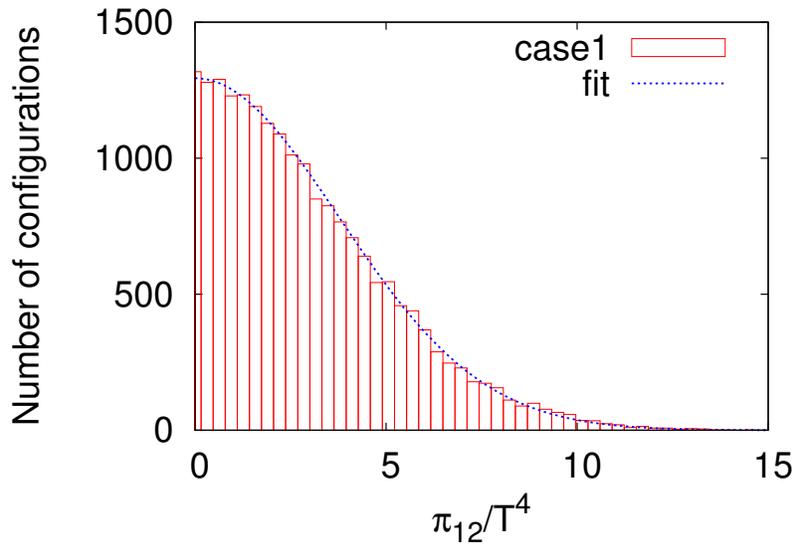}
\end{center}
\caption{Distribution of fluctuation of $\bar{\pi}_{12}$ obtained from numerical simulations for case1.
The horizontal axis is normalized by the temperature.
The dotted line denotes the Gaussian fit.}
\label{fig:pi}
\end{figure}

\begin{figure}
\begin{center}
\includegraphics*[angle=0,scale=0.9]{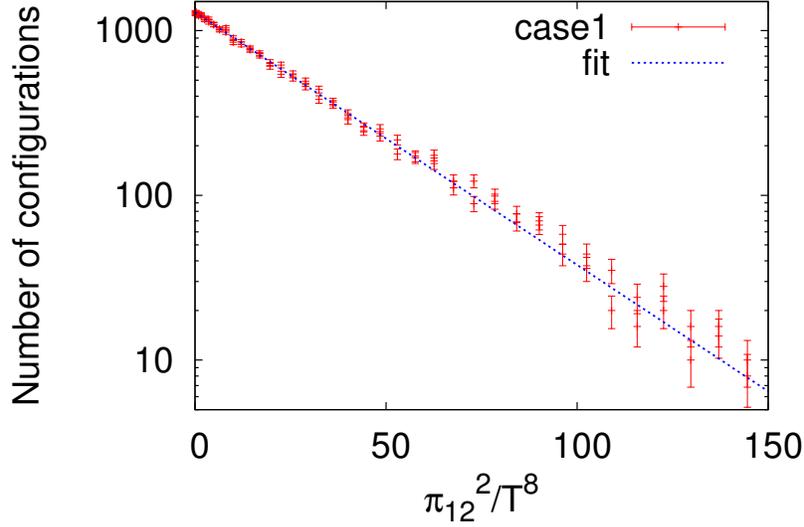}
\end{center}
\caption{Distribution of ${\pi_{12}}^2$ obtained from numerical simulations for case 1.
The vertical axis is logarithmic.
$\beta_2$ can be read from the gradient of fit line.}
\label{fig:pi2}
\end{figure}

In this work, we perform the lattice simulation for
SU(3) pure gauge theory with the standard Wilson gauge action.
Configurations are generated by heatbath and overrelaxation
algorithms.
In Table~\ref{tbl:ltx-tbl2} we show lattice parameters used 
in this work.
The simulation is performed with three lattice sizes 
$N_\sigma^3\times N_\tau$ with a periodic boundary condition 
and different lattice spacing $a$.
For each parameter, 10000 configurations have been prepared.
The energy momentum tensor is given by
$T^{\mu\nu} = 2{\rm tr} \left[ F^\mu_\rho F^{\rho\nu} 
+ (1/4) g^{\mu\nu} F^{\rho\sigma}F_{\rho\sigma} \right]$.
For the definition of the field strength on the lattice,
we have chosen the clover operator.

For the spatial average in Eq.~(\ref{bar}),
we take the average of $T_{\mu\nu}$ on a cube of size 
$(N_\sigma/2)^3$ to remove the effect of periodicity.
For each configuration one can take eight such subconfigurations
on cubes. We regard two of the cubes aligned in a diagonal direction as
statistically independent areas.
Furthermore, we adopt two sets of subconfigurations on the two time
slices at $N_{\tau} = 0$ and $N_{\tau} = 1/2aT$ in order to
improve statistics. 
In this way, four subconfigurations are taken from each 
configuration for analysis.

In Fig.~\ref{fig:pi}, we show the distribution of the spatial 
average of the shear viscous pressure $\bar\pi_{12}=\bar T_{12}$ 
for case 1.
The same result is shown in Fig.~\ref{fig:pi2} with different
choices of vertical and horizontal axes.
These figures show that the distribution of $\bar\pi_{12}$ 
is of Gaussian form, as anticipated.
Extracting fluctuations from this distribution and substituting
the result to Eq.~(\ref{pi^2}), we obtain 
\begin{equation}
\beta_2 \simeq 0.0036. 
\label{rb2}
\end{equation}

This value of $\beta_2$ is, however, quite small
in the sense that the relaxation time $\tau_\pi = 2\beta_2\eta$ is
so small for  $\eta\lesssim 1$ that IS theory turns back to the
first order theory. 
In fact, the speed of transverse plane wave, $v_T$, with 
the value of $\beta_2$ in Eq.~(\ref{rb2}) gives \cite{AM2}
\begin{equation}
\frac{v_T}{c} = \sqrt{\frac{1}{2\beta_2 (\epsilon +p)}} 
\simeq 6,
\label{vt}
\end{equation}
where we have used the values of $\epsilon$ and $p$ obtained
in the previous work on bulk thermodynamics of SU(3) gauge theory
\cite{MO}.
Eq.~(\ref{vt}) means that causality is violated with $\beta_2$ obtained in this
analysis.
Similar results are obtained on cases 2 and 3.

\section{Discussions}

In this work, we tried to evaluate the ratio between the relaxation 
time $\tau_\pi$ and the shear viscosity $\eta$ in SU(3) gauge theory
by lattice simulations.
Using BE principle and the form of entropy in IS theory, the ratio is
related to statistical fluctuation of stress tensor in equilibrium 
through Eq.~(\ref{pi^2}).
In the numerical simulations, however, we have observed large fluctuations
of stress tensor, $\langle \bar\pi_{12}^2 \rangle$, which leads to
unacceptably small $\beta_2$ violating causality.

It is, however, suspicious to simply apply fluctuations obtained on
the lattice to Eq.~(\ref{pi^2}), since $\langle \bar\pi_{12}^2 \rangle$ 
is actually an ultraviolet divergent quantity and it diverges
even in the vacuum. The large fluctuations would have originated from
this vacuum fluctuations, which should be subtracted to obtain
the physical value of fluctuations.
This analysis is in progress. 
The present numerical results, however, indicate that we need more 
statistics to separate physical fluctuations from the vacuum one 
with a good accuracy.

\section*{Acknowledgement}
The authors thank S. Pratt for valuable discussions and comments.
This work is supported by the Large Scale Simulation Program No.~09-11
(FY2009) of High Energy Accelerator Research Organization (KEK).

\end{document}